\documentclass[11pt]{article}
\usepackage{amsmath, amssymb, amscd, amsthm, amsfonts}
\usepackage{graphicx}
\usepackage{hyperref}
\usepackage{braket}
\usepackage{authblk}
\usepackage{cite}

\title{Evaluation of Femtosecond Time-bin Qubits Using Frequency Up-conversion Technique}

\author[1, 2]{Yuta Kochi}
\author[3]{Sunao Kurimura}
\author[1, 2]{Junko Ishi-Hayase}

\affil[1]{\footnotesize School of Fundamental Science and Technology,  Keio University, Kanagawa 223-8522, Japan}
\affil[2]{Center for Spintronics Research Network,  Keio University, Kanagawa 223-8522, Japan}
\affil[3]{National Institute for Materials Science, Ibaraki 305-0047, Japan}
\date{}

\begin{document}

\maketitle

\begin{abstract}
Time-bin qubits, in which quantum information is encoded in a single photon at different times $\Delta t$, are suitable for long-distance transmission via optical fibers. However, detection of time-bin qubits has been limited to the nanosecond range owing to the limited temporal resolution of single-photon detectors. In this study, we developed an up-conversion single-photon detector (UCSPD), using commercial nonlinear crystals of different lengths. By changing the crystal length and pump power, we quantitatively evaluated the efficiency and temporal resolution of the UCSPD and determined the optimal conditions for measuring femtosecond time-bin qubits. This detector achieved a temporal resolution of 415 fs and up-conversion efficiency of 10.1 \%. Consequently, we successfully evaluated single-photon level pseudo femtosecond time-bin qubits with a pulse interval of only 800 fs.
\end{abstract}

\section{Introduction}\label{section-introduction}
Recently, quantum information processing and communication using photons have been attracting attention. Time-bin qubits, in which quantum information is encoded in a single photon at different times, are well suited for long-distance transmission via optical fibers. The storage of time-bin qubits in quantum memory has also been demonstrated using entangled photons\cite{Martin2013,Saglamyurek2015,Lee2018,Liu2020}. The relative phase $\phi$ of the time-bin qubits can be evaluated by measuring the interference between the early and late pulses with a pulse interval $\Delta t$ using an interferometer. These interfered time-bin qubits consist of three pulses and the center pulse amplitude changes depending on the relative phase $\phi$. Hence, by selectively detecting the center pulse, we can evaluate the relative phase $\phi$ of the time-bin qubits (as shown in Fig. 1(a)). To increase the temporal density of the qubits, it is necessary to reduce both the pulse width and pulse interval $\Delta t$ of the time-bin qubits. However, the pulse interval $\Delta t$ is limited to the order of nanoseconds owing to the temporal resolution of single-photon detectors; although a single-photon pulse with a 38.9 fs FWHM has already been generated in the telecommunication wavelength band\cite{Zhu2013}. Focusing on the relative phase evaluation, a time-bin qubit evaluation with a pulse interval $\Delta t$ of 2.2 ps was performed by Donohue \textit{et al.}\cite{Donohue2013}. They obtained phase information from the changes in the frequency components. Although the measurable pulse interval $\Delta t$ depended on the resolution of the monochromator, the considered limit was $\Delta t\sim 1$ ps.

\begin{figure}[htbp]
  \centering
  \includegraphics[width=13cm]{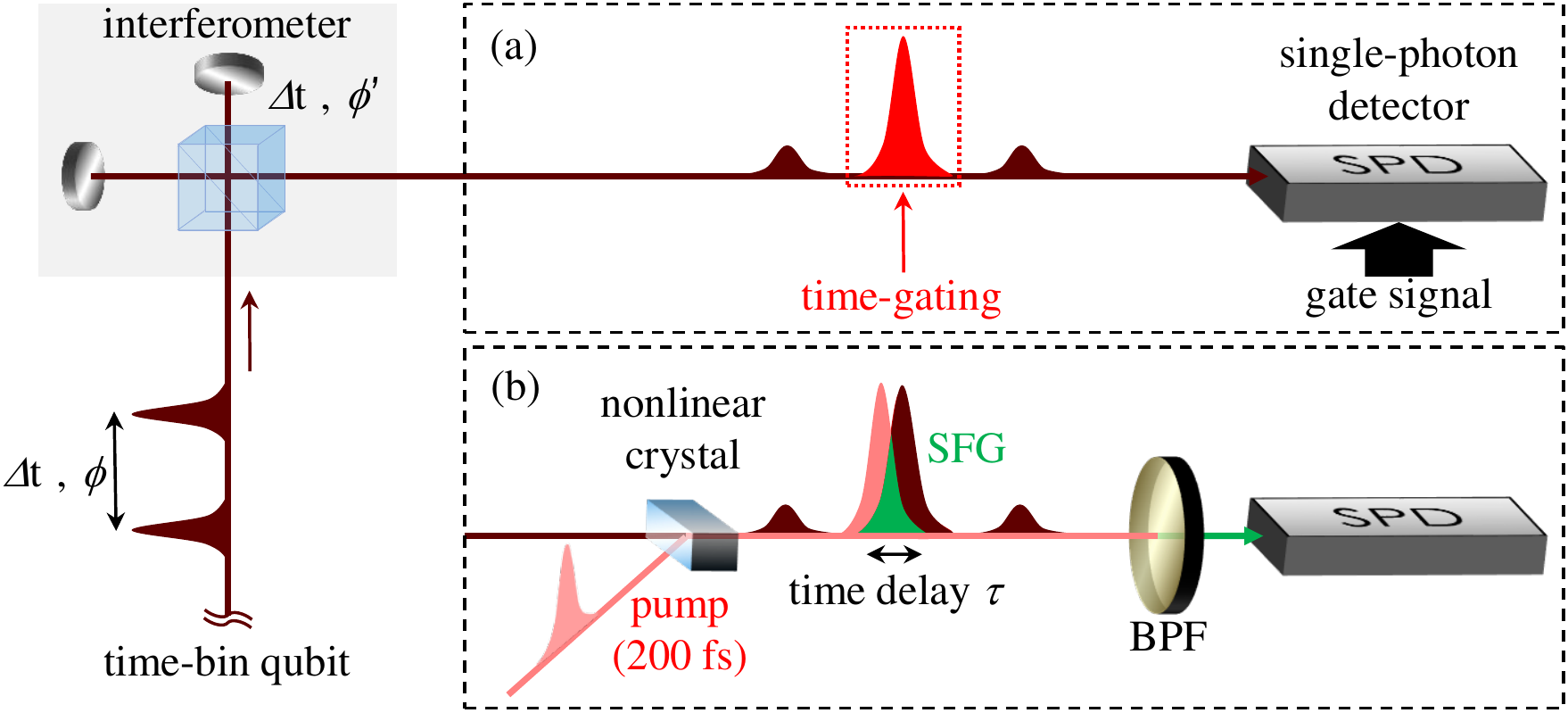}
\caption{(a) Previous method of evaluating time-bin qubits and (b) the current method using frequency up-conversion technique. SFG: sum-frequency generation. BPF: bandpass filter.}
\end{figure}

To improve temporal resolution, frequency up-conversion single-photon detectors (UCSPD) have been developed. These detectors were initially developed to convert telecommunication wavelength photons to shorter wavelengths that can be more efficiently detected by commercial single-photon detectors such as Si avalanche photodiodes (APD) instead of InGaAs APDs\cite{Midwinter1967}. Currently, UCSPDs that use a CW laser as a pump light, achieve conversion efficiencies of more than 90 \%\cite{VanDevender2003, Albota2004, Yao2020}. In these types of UCSPDs, the temporal resolution and dead time depends on the performance of the Si APDs. However, femtosecond order temporal resolutions have been achieved using a femtosecond pulse laser as a pump light for UCSPDs. Kuzucu \textit{et al.}\cite{Kuzucu2008} measured the correlation of the time waveform of photon pairs using a UCSPD with a 1-mm-long nonlinear crystal and a femtosecond pulse laser as the pump light. Allgaier \textit{et al.}\cite{Allgaier2017, Allgaier2017b} also achieved a high temporal resolution and high conversion efficiency using a 27-mm-long nonlinear crystal with a waveguide that satisfied the group velocity matching (GVM) between the pump light and signal light.
However, there have been no reports of the application of UCSPD to detect femtosecond time-bin qubits. Moreover, the quantitative evaluation of the crystal length dependence of the temporal resolution and conversion efficiency in the up-conversion process has not yet been investigated.

In this study, we developed a UCSPD to evaluate femtosecond time-bin qubits (Fig. 1(b)). For this detector, we used commercial nonlinear crystals of different lengths. By changing the crystal length and pump power, we quantitatively evaluated the efficiency and temporal resolution of the UCSPD and determined the optimal conditions for measuring femtosecond time-bin qubits. Consequently, we successfully evaluated the relative phase of the shortest (pulse interval $\Delta$t = 800 fs) pseudo femtosecond time-bin qubits ever reported.

\section{Performance of UCSPD}
\subsection{Experimental setup}
To enable measurements that exceed the performance of conventional single-photon detectors, we developed a UCSPD. This single-photon detector uses the frequency up-conversion technique, and when signal photons are inserted into this detector, they are converted to visible photons by sum-frequency generation (SFG). In this process, when we use a femtosecond pulse laser as the pump light, SFG occurs only when the signal photons and the pump pulse overlap. Therefore, to detect the frequency component of SFG with commercial single-photon detectors such as Si APDs, we can time-gate the signal waveforms with femtosecond order temporal resolution.

Figure 2 shows the schematic setup of the UCSPD. We used an 820 nm femtosecond Ti:Sapphire laser (Coherent Mira-XP, pulse width $\Delta_{\rm{p}}$: 200 fs, repetition frequency: 76.3 MHz) to pump nonlinear crystals, and 1520 nm pulse from optical parametric oscillator (Coherent Mira OPO-X, pulse width $\Delta_{\rm{s}}$: 240 fs) as a signal light. We selected periodically poled 1 mol.\% MgO-doped stoichiometric lithium tantalate bulk crystals (PPMgSLT) for frequency up-conversion. The crystal lengths $L$ were 1, 2, and 3-mm-long, and the periodically poled periods were 8.55 $\mu$m. This period is suitable for type-0 phase-matched SFG ($820\ \mathrm{nm} + 1520\ \mathrm{nm} \rightarrow 533\ \mathrm{nm}$) at a crystal temperature of 85$^\circ$C. The pump and signal pulses are focused by different lenses to prevent a focal position shift owing to chromatic aberration. Finally, the up-converted photons were detected by Si APD (Hamamatsu C13001-01).
\begin{figure}[htbp]
  \centering
  \includegraphics[width=15cm]{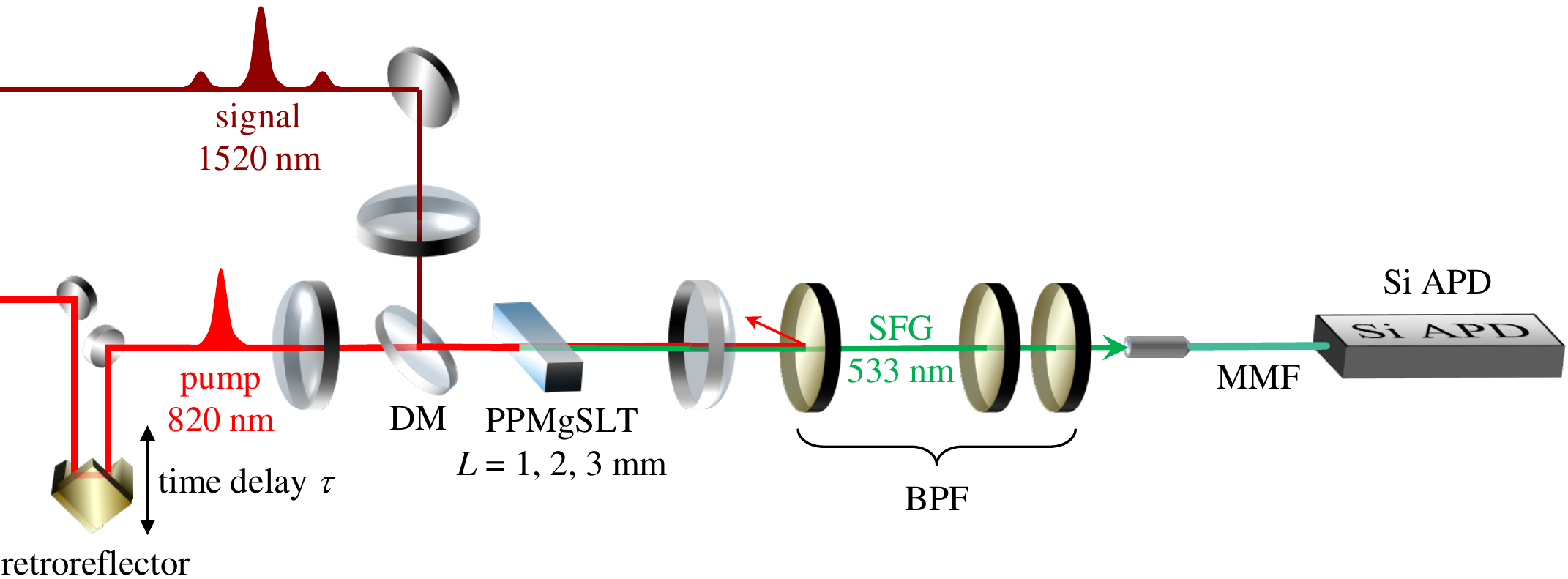}
\caption{Optical system of the UCSPD. BPF: band pass filter. DM: dichroic mirror. MMF: multimode fiber. Si APD: silicon avalanche photodiode.}
\end{figure}

\subsection{Results and discussion}
First, we measured a single pulse at the single-photon level (average photon number: 0.1/pulse) using the UCSPD to estimate the temporal resolution and up-conversion efficiency. The pump power was set to 300 mW. When we changed the PPMgSLT crystals to different lengths, we moved each lens to readjust the pump and signal light focal positions to optimize up-conversion efficiency. Figure 3(a) shows the count rate of the up-converted photons measured as a function of the time delay between the signal and pump pulse. The peak count increases when the crystal length $L$ is changed from 1 to 2 mm but saturates for the crystal with $L=3$ mm. This is because the temporal overlap between the pump pulse and signal pulse decreases owing to the group velocity dispersion in the crystal. To estimate the temporal resolution, we define the temporal resolution function $T(t)$, which can be described as:
\begin{equation}
    \begin{split}
        T(t)&\equiv P(t)*G(t)\\
        G(t)&=
         \begin{cases}
            1&-\frac{\tau_{\rm g}L}{2}\leq t\leq\frac{\tau_{\rm g}L}{2}\\
            0&t<-\frac{\tau_{\rm g}L}{2},\ \frac{\tau_{\rm g}L}{2}<t,
         \end{cases}
    \end{split}
\end{equation}
where $P(t)$ is the temporal waveform of the pump pulse and $G(t)$ is a rectangular function with a width of group delay $\tau_g L$ (in our experiment, $\tau_g$ is 204.3 fs/mm\cite{Lim2013}) between the pump pulse and signal pulse. Using this function $T(t)$ and the signal waveform $S_{\rm{in}}(t)$, the waveforms measured by the UCSPD $S_{\rm{out}}(t)$ were calculated as:
\begin{equation}
    S_{\rm{out}}(t)=S_{\rm{in}}(t)*T(t).
\end{equation}
Assuming that the temporal waveforms of the pump and signal pulses are Gaussian, 
\begin{equation}
    \begin{split}
        P(t)&\propto\exp\left[-\left(\frac{2\sqrt{\ln2}}{\Delta_{\rm p}}t\right)^2\right]\\
        S_{\rm{in}}(t)&\propto\exp\left[-\left(\frac{2\sqrt{\ln2}}{\Delta_{\rm s}}t\right)^2\right],
    \end{split}
\end{equation}
the waveform $S_{\rm{out}}(t)$ can be calculated as:
\begin{equation}
\label{temp}
    \begin{split}
        S_{\rm{out}}(t)&=S_{\rm{in}}(t)*T(t)\\
        &\propto\int_{-\tau_{\rm g}L/2}^{\tau_{\rm g}L/2}\exp\left[-\frac{4\ln2}{\Delta_{\rm p}^2+\Delta_{\rm s}^2}\left(t-\tau\right)^2\right]d\tau\\
        &=\int_{-\tau_{\rm g}L/2-t}^{\tau_{\rm g}L/2-t}\exp\left[-\frac{4\ln2}{\Delta_{\rm p}^2+\Delta_{\rm s}^2}\tau'^2\right]d\tau'\\
        &\propto\rm{erf}\left[\sqrt{\frac{\ln2}{\Delta_{\rm p}^2+\Delta_{\rm s}^2}}\left(\tau_{\rm{g}}\it L-2t\right)\right]-\rm{erf}\left[\sqrt{\frac{\ln2}{\Delta_{\rm p}^2+\Delta_{\rm s}^2}}\left(-\tau_{\rm{g}}\it L-2t\right)\right].
    \end{split}
\end{equation}
This equation expresses that when the crystal length $L$ increases, the group delay $\tau_{\rm g}L$ also increases and the measured waveform $S_{\rm{out}}(t)$ approaches a rectangular waveform. In this study, we define the temporal resolution as FWHM of $T(t)$. From the experimental results in Fig. 3(a), we estimated that the temporal resolutions of the UCSPD were 255, 415, and 591 fs for crystal length $L$ of 1, 2, and 3-mm-long, respectively. The solid lines in Fig. 3(a) and (b) were calculated using Eq. (4), and it can be observed that they are well consistent with the experimental results.
\begin{figure}[htbp]
  \centering
  \includegraphics[width=14cm]{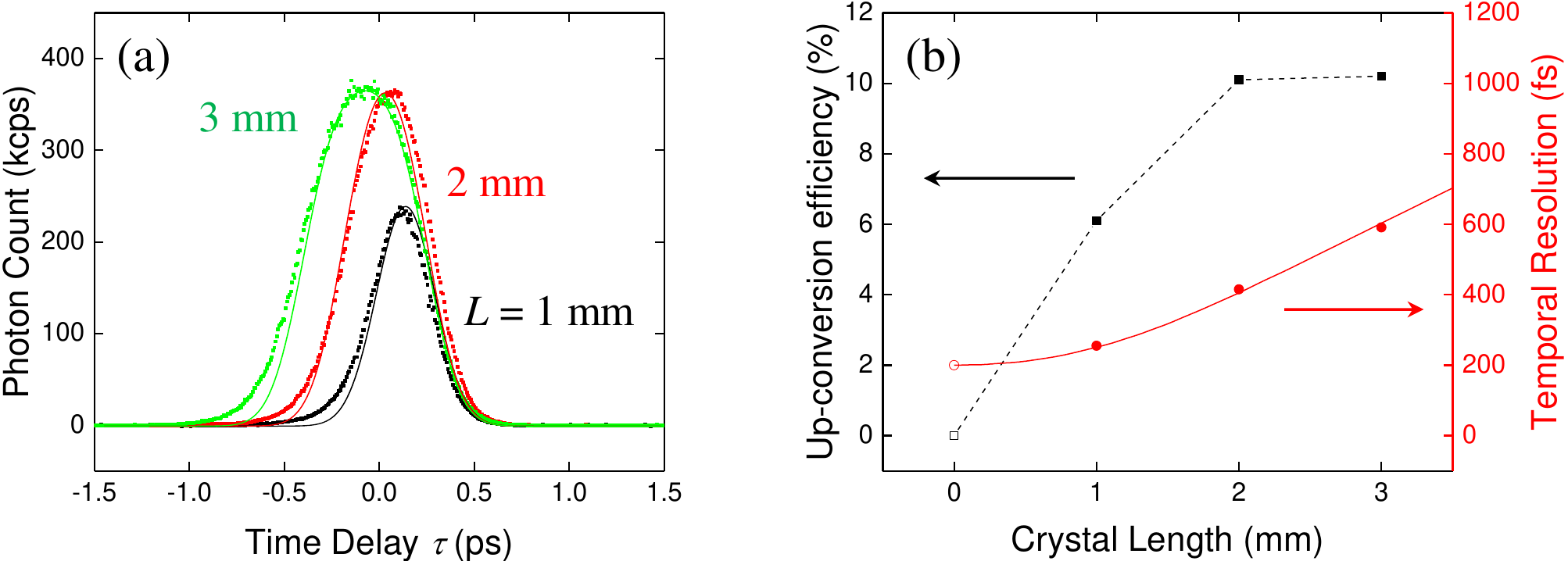}
\caption{(a) Experimental and theoretical measurement waveforms by the UCSPD. (b) Temporal resolution (black) and up-conversion efficiency (red) dependence of the crystal length. The square dots represent the measurement results. The solid line in each figure is calculated using Eq. (4).}
\end{figure}

Next, we studied the pump power dependence of the up-conversion efficiency $\eta_{\rm{UCSPD}}$ and the detection limit of the UCSPD. Figure 4(a) shows that the up-conversion efficiency increases as the pump power increases, but the slope gradually decreases. However, the noise count rate increases exponentially as the up-conversion efficiency increases (Fig. 4(b)). Considering that the dark count of the Si APD is 100 cps or less, these noises are mainly photons from the PPMgSLT crystals. The main noise photons are generated by the spontaneous parametric down-conversion (SPDC) process and reconverted by the pump light, as the number of photons produced by Raman scattering is sufficiently small in a few millimeters of crystal\cite{Kamada2008}. It has been reported that noise derived from SPDC can be avoided by setting the pump wavelength to a long wavelength of approximately 1.9 $\mu$m\cite{Shentu2013, Liang2019, Zhang2020}.

\begin{figure}[htbp]
  \centering
  \includegraphics[width=13cm]{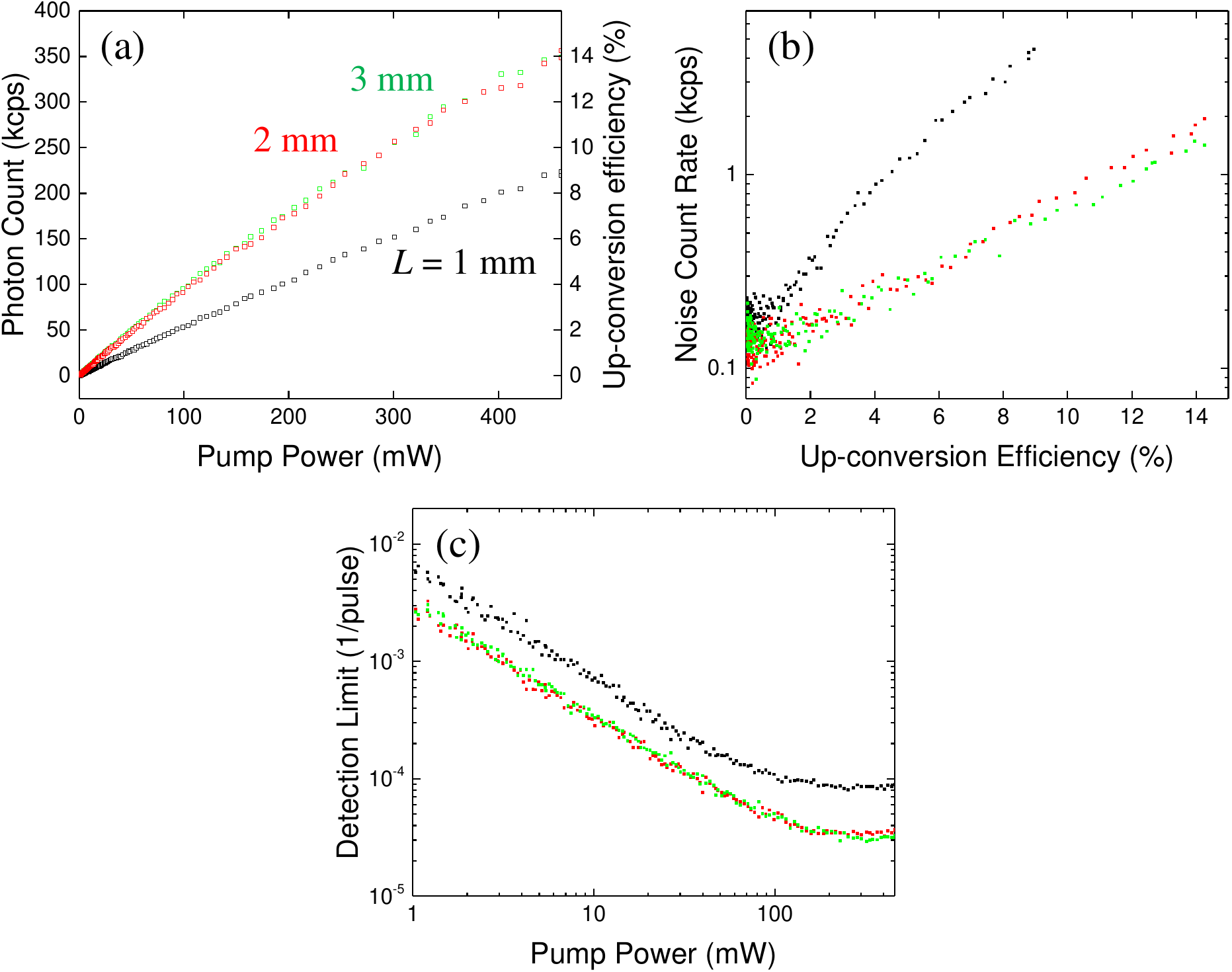}
\caption{(a) Pump power dependence of the up-conversion efficiency. black: 1 mm. red: 2 mm. green: 3 mm long crystal. (b) Change in noise count rate. (c) Average number of photons per pulse at the detection limit.}
\end{figure}

Subsequently, we defined the number of photons that can detect three times the standard deviation of the noise count rate $\sigma_{\rm{NCR}}$ as the detection limit:
\begin{equation}
    \left(\rm{the\ average\ photon\ number\ of\ the\ detection\ limit}\right)=\frac{3\sigma_{\rm{NCR}}}{\eta_{\rm{UCSPD}}}.
\end{equation}
In Fig. 4(c), we found that the detection limit was almost constant even with a pump power of 200 mW or higher, and it acquired minimum values of 8.6, 3.3, and 3.1 $\times10^{-5}$ /pulse ($L$ was 1, 2, and 3-mm-long, respectively) when the pump power was 300 mW. Considering the detection efficiency of the Si APD: 41 \%, the fiber coupling efficiency: 85 \%, and the transmittance of the filters: 94 \%, we estimated the internal up-conversion efficiency of the UCSPD as 6.1, 10.1, and 10.2 \% ($L$ was 1, 2, and 3-mm-long, respectively) when the pump power was 300 mW (Table 1). Comparing our UCSPD with other detectors, we achieved less noise than InGaAs at room temperature (300 K) and a high time resolution on the order of femtoseconds (Table 2). Allgaier's group of studies \cite{Allgaier2017, Allgaier2017b} achieved a high up-conversion efficiency and temporal resolution satisfying GVM between the pump and signal pulse using type-II phase matching. However, to satisfy this GVM, they used a PPLN crystal with a periodically poled period of 4.4 $\mu$m and a crystal length of 27 mm with an optical waveguide, which is difficult to obtain and produce. In contrast, our UCSPD achieved a femtosecond order temporal resolution and practical detection efficiency using readily available commercial crystals.
\begin{table}[ht]
 \caption{UCSPD performance.}
 \centering
  \begin{tabular}{rrrr}
   \hline
    \multicolumn{1}{c}{Crystal} & \multicolumn{1}{c}{Temporal}& \multicolumn{1}{c}{Up-conversion} &\multicolumn{1}{c}{Noise} \\
    \multicolumn{1}{c}{Length} & \multicolumn{1}{c}{Resolution} & \multicolumn{1}{c}{Efficiency} &\multicolumn{1}{c}{Count Rate}  \\
   \hline\hline
   1 mm & 255 fs & 6.1 \% & 1910 cps\\
   2 mm & 415 fs & 10.1 \% & 800 cps\\
   3 mm & 591 fs & 10.2 \% & 700 cps\\
   \hline
  \end{tabular}
\end{table}

\begin{table}[ht]
 \caption{Comparison with other single photon detectors.}
 \centering
  \begin{tabular}{crrrrr}
   \hline
    Single Photon & \multicolumn{1}{c}{Operating} & \multicolumn{1}{c}{Temporal} & \multicolumn{1}{c}{Quantum} & \multicolumn{1}{c}{Noise} & \multicolumn{1}{c}{Dead}\\
    Detector&\multicolumn{1}{c}{Temperature} & \multicolumn{1}{c}{Resolution} & \multicolumn{1}{c}{Efficiency} & \multicolumn{1}{c}{Count Rate}& \multicolumn{1}{c}{Time}\\
   \hline\hline
    InGaAs/InP\cite{Jiang2017} & 223 K & 170 ps & 27.5 \%& 1200 cps& 100 ns\\
    SNSPD & 2.1 K& 57 ps & 80 \% & 10 cps & 10 ns\\
    \multicolumn{1}{l}{\ \ UCSPD (CW)\cite{Yao2020}} & 300 K& 400 ps & 40.2 \% & 200 cps & 100 ns\\
    \multicolumn{1}{l}{\ \ UCSPD (pulse)\cite{Allgaier2017, Allgaier2017b}} & 300 K& 0.3 ps & 16.9 \%& & \multicolumn{1}{c}{-}\\
    \multicolumn{1}{l}{\ \ UCSPD (this work)} & 300 K & 0.4 ps & 3.3 \%& 700 cps & \multicolumn{1}{c}{-}\\
   \hline
  \end{tabular}
\end{table}

Using these experimental results, we set optimum measurement conditions as: a crystal length of 2 mm, pump power of 300 mW (in these conditions, temporal resolution of 415 fs, up-conversion efficiency of 10.1 \%, and average photon number of the detection limit as $3.3\times10^{-5}$ /pulse).

\section{Evaluation of femtosecond time-bin qubits}
\subsection{Experimental setup}
Figure 5 shows the experimental setup for creating and evaluating pseudo femtosecond time-bin qubits. This setup consists of two unbalanced Michelson interferometers, each of which has the same time delay $\Delta t$ that determines the pulse interval of the time-bin qubits. These interferometers have one polarization beam splitter and two quarter-wave plates to reduce the loss of returning light. A cover glass is inserted into one arm of the interferometer, and any relative phase can be obtained by changing the angle of the glass plate. In this experiment, we used a single-photon level pulse (average photon number: 0.1 /pulse) from an optical parametric oscillator as the light source. We set $\Delta t$ to 800 fs which is significantly larger than the pulse width and temporal resolution of the UCSPD. When a single photon $ \ ket{1}$ passes through the interferometer, it becomes a time-bin qubit encoded as $\ket{1}\rightarrow\frac{1}{\sqrt2}(\ket{1}+e^{i\phi}\ket{2})$, where $\ket{1}$ is the photon through the short arm and $\ket{2}$ passes through the long arm in the interferometer. Subsequently, when the time-bin qubits pass through the interferometer again, each photon state becomes $\ket{n}\rightarrow\frac{1}{\sqrt2}(\ket{n}+e^{i\phi'}\ket{n+1})$, and the output state $\ket{\psi_{\rm{out}}}$ of the interferometer can be represented as:
\begin{equation}
\ket{\psi_{\rm{out}}}\propto\ket{1}+(e^{i\phi}+e^{i\phi'})\ket{2}+\ket{3}.
\end{equation}
In this situation, the ratio of the probability amplitudes $P_1$, $P_2$, and $P_3$ (photon states are $\ket{1}$, $\ket{2}$, and $\ket{3}$, respectively) is:
\begin{equation}
P_1:P_2:P_3=1:4\cos^{2}\left(\frac{\phi-\phi'}{2}\right):1.
\end{equation}
Therefore, by time-gating the center pulse $P_2$ using the UCSPD, we can evaluate the relative phase $\phi-\phi'$ of time-bin qubits.
\begin{figure}[htbp]
  \centering
  \includegraphics[width=13cm]{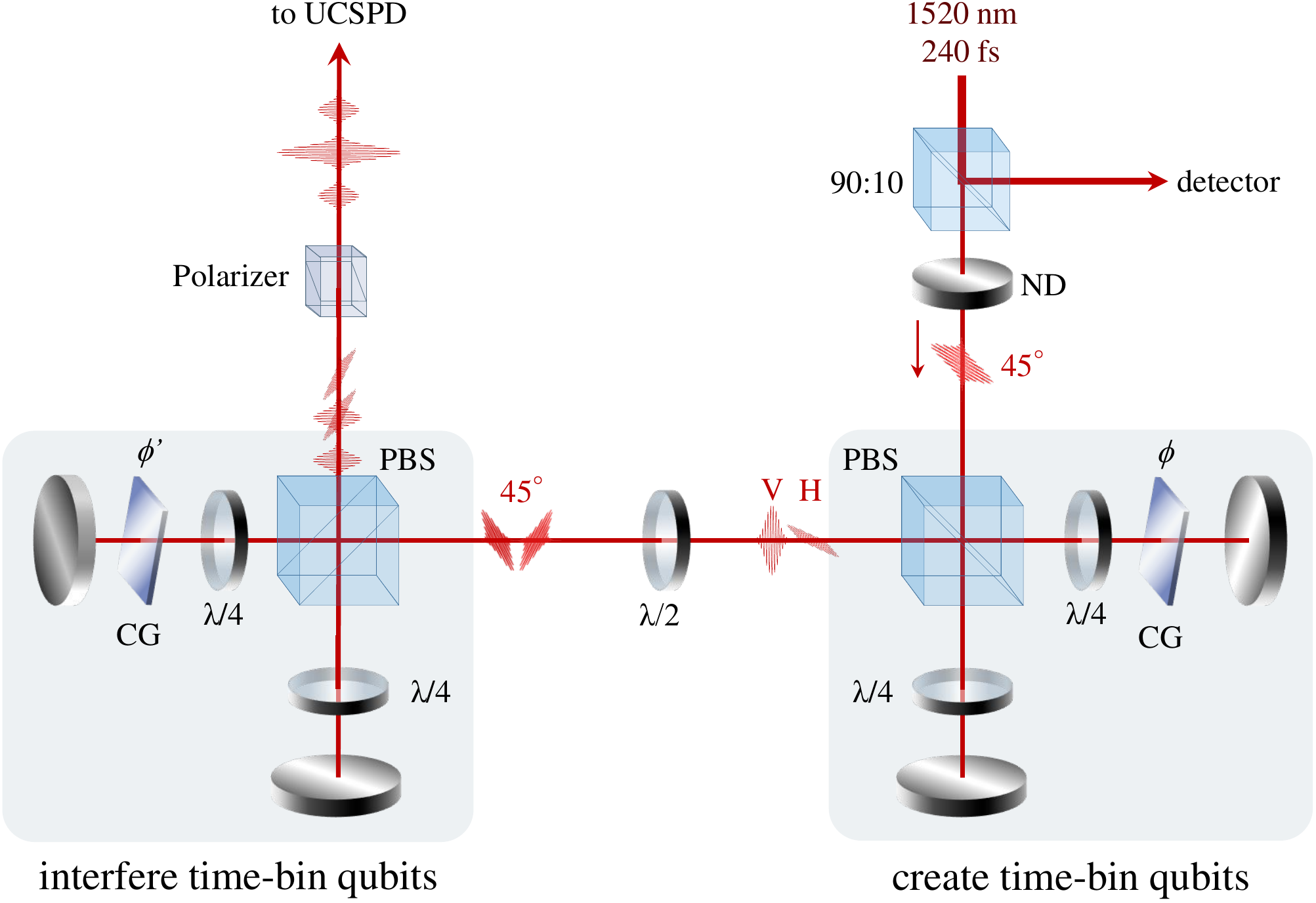}
\caption{Experimental setup of the creation and evaluation of femtosecond time-bin qubits. CG: cover glass. ND: neutral density filter. PBS: polarization beam splitter. $\lambda/2$: half-wave plate. $\lambda/4$: quarter-wave plate.}
\end{figure}

\subsection{Results and discussion}
Figure 6(a) shows the measured temporal waveforms of the pseudo femtosecond time-bin qubits with different relative phases. Time-bin qubits interference waveforms with a pulse interval of only 800 fs were clearly observed at the position where the delay time of the pump pulse was 0 fs. The solid blue line is an estimation of the original waveforms by deconvolution using the temporal resolution function $T(t)$, and it was found that the original waveform can be reconstructed from the measurement results.
\begin{figure}[htbp]
  \centering
  \includegraphics[width=12cm]{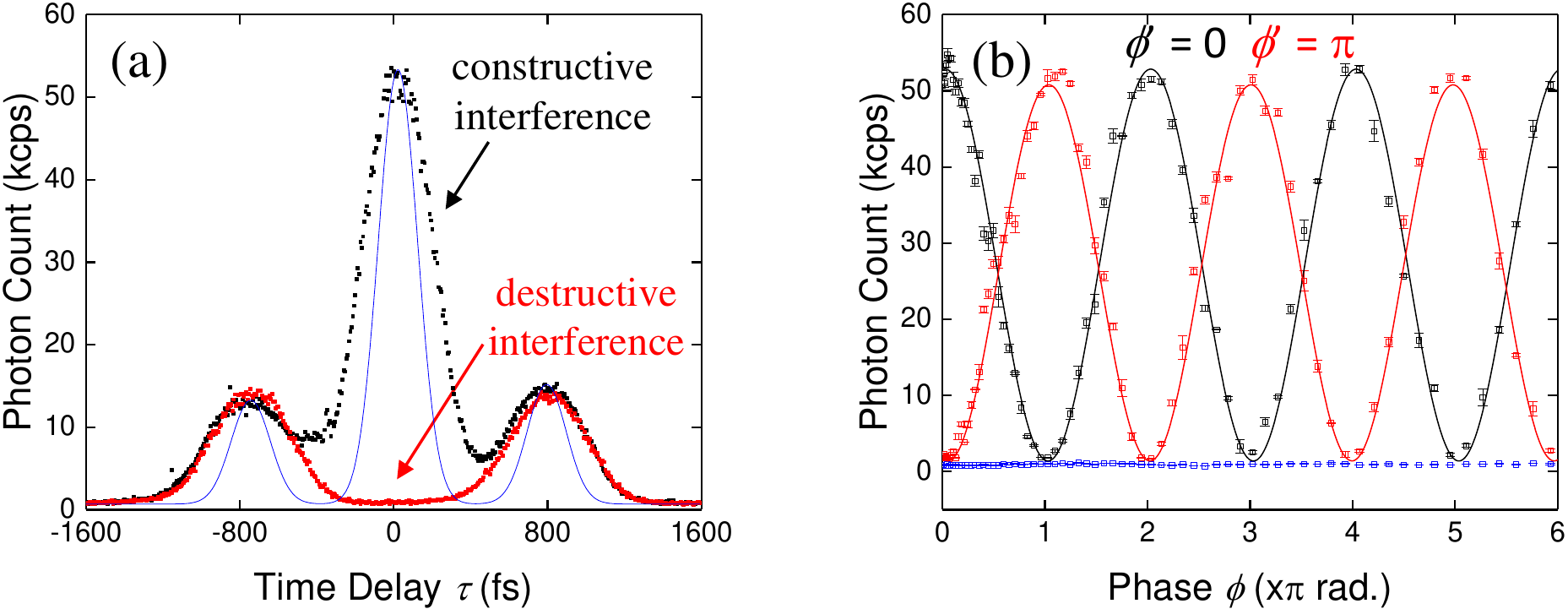}
\caption{(a) Temporal waveforms of time-bin qubits. The black and red dots represent the experimental results, and the solid blue line is an estimation of the original waveforms by deconvolution using the temporal resolution function $T(t)$. (b) Count change of the center peak at $\tau$ is 0 fs in (a). The solid lines are sine curve fitting of the experimental results. black: $\phi'=0$. red: $\phi'=\pi$.}
\end{figure}

Next, we fixed the delay time as 0 fs to measure the change in the interference intensity $P_2$ owing to the change in the relative phase $\phi-\phi'$. Figure 6(b) shows the count change when the relative phase is changed, and the visibility is $98.2\pm0.09$ and $98.1\pm0.09$ \% ($\phi$' = 0, $\pi$, respectively).

The evaluation of picosecond time-bin qubits ($\Delta t$ = 2.2 ps) was studied by Donohue \textit{et al.} in 2013\cite{Donohue2013}. They used the frequency up-conversion technique with a positive chirped signal and negatively chirped pump pulses, and then evaluated them in the frequency domain. The up-conversion efficiency was 0.06 \%, and the detection efficiency of the monochromator was 6 \%. In their method, the pulse interval $\Delta t$ of the time-bin qubits depends on the resolution of the monochromator, with a 1 ps limit. Moreover, improving the resolution of the monochromator further reduces the detection efficiency.
In contrast, in our method, the temporal resolution depends on the pulse width and group velocity dispersion of the PPMgSLT crystals. Regarding the group velocity dispersion, it has been demonstrated that GVM can be satisfied by using type-I quasi-phase matching in PPLN crystals between the wavelengths of 820 and 1520 nm, and a further improvement in the up-conversion efficiency and the temporal resolution is expected\cite{Yu2002, Yu2003, Allgaier2017, Allgaier2017b}.

\section{Conclusion}
In this study, we developed a UCSPD using commercial PPMgSLT crystals. We estimated the temporal resolution of the UCSPD to be 415 fs, which cannot be achieved with conventional electronic single-photon detectors, and the up-conversion efficiency to be 10.1 \%. Moreover, we successfully evaluated single-photon level femtosecond time-bin qubits with a pulse interval $\Delta t$ of 800 fs which is the shortest reported $\Delta t$. By applying this method to quantum communication, it is expected that the information density will be increased by three orders of magnitude and it will further speed up communication. In future, we plan to evaluate time-bin qubits with nonclassical light and store/reproduce them in broadband quantum memory using quantum dots.

\section*{Acknowledgments}
This work was supported by MEXT Q-LEAP (No. JPMXS0118067395), and CSRN, Keio University.

\bibliographystyle{unsrt}
\bibliography{references} % see references.bib for bibliography management

\begin{thebibliography}{10}

\bibitem{Martin2013}
A.~Martin, F.~Kaiser, A.~Vernier, A.~Beveratos, V.~Scarani, and S.~Tanzilli.
\newblock {Cross time-bin photonic entanglement for quantum key distribution}.
\newblock {\em Physical Review A - Atomic, Molecular, and Optical Physics},
  87(2):1--5, 2013.

\bibitem{Saglamyurek2015}
Erhan Saglamyurek, Jeongwan Jin, Varun~B. Verma, Matthew~D. Shaw, Francesco
  Marsili, Sae~Woo Nam, Daniel Oblak, and Wolfgang Tittel.
\newblock {Quantum storage of entangled telecom-wavelength photons in an
  erbium-doped optical fibre}.
\newblock {\em Nature Photonics}, 9(2):83--87, 2015.

\bibitem{Lee2018}
J.~P. Lee, L.~M. Wells, B.~Villa, S.~Kalliakos, R.~M. Stevenson, D.~J.P. Ellis,
  I.~Farrer, D.~A. Ritchie, A.~J. Bennett, and A.~J. Shields.
\newblock {Controllable Photonic Time-Bin Qubits from a Quantum Dot}.
\newblock {\em Physical Review X}, 8(2):21078, 2018.

\bibitem{Liu2020}
Chao Liu, Tian~Xiang Zhu, Ming~Xu Su, You~Zhi Ma, Zong~Quan Zhou, Chuan~Feng
  Li, and Guang~Can Guo.
\newblock {On-Demand Quantum Storage of Photonic Qubits in an On-Chip
  Waveguide}.
\newblock {\em Physical Review Letters}, 125(26):260504, 2020.

\bibitem{Zhu2013}
E.~Y. Zhu, Z.~Tang, L.~Qian, L.~G. Helt, M.~Liscidini, J.~E. Sipe, C.~Corbari,
  A.~Canagasabey, M.~Ibsen, and P.~G. Kazansky.
\newblock {Poled-fiber source of broadband polarization-entangled photon
  pairs}.
\newblock {\em Optics Letters}, 38(21):4397, 2013.

\bibitem{Donohue2013}
John~M. Donohue, Megan Agnew, Jonathan Lavoie, and Kevin~J. Resch.
\newblock {Coherent ultrafast measurement of time-bin encoded photons}.
\newblock {\em Physical Review Letters}, 111(15):1--5, 2013.

\bibitem{Midwinter1967}
J.~E. Midwinter and J.~Warner.
\newblock {Up-conversion of near infrared to visible radiation in
  lithium-meta-niobate}.
\newblock {\em Journal of Applied Physics}, 38(2):519--523, 1967.

\bibitem{VanDevender2003}
Aaron~P. VanDevender and Paul~G. Kwiat.
\newblock {High-efficiency single photon detection via frequency
  up-conversion}.
\newblock {\em Quantum Information and Computation}, 5105(August 2003):216,
  2003.

\bibitem{Albota2004}
Marius~A. Albota and Franco N.~C. Wong.
\newblock {Efficient single-photon counting at 1.55 µm by means of frequency
  upconversion}.
\newblock {\em Optics Letters}, 29(13):1449, 2004.

\bibitem{Yao2020}
Ni~Yao, Quan Yao, Xiu-Ping Xie, Yang Liu, Peizhen Xu, Wei Fang, Ming-Yang
  Zheng, Jingyun Fan, Qiang Zhang, Limin Tong, and Jian-Wei Pan.
\newblock {Optimizing up-conversion single-photon detectors for quantum key
  distribution}.
\newblock {\em Optics Express}, 28(17):25123, 2020.

\bibitem{Kuzucu2008}
Onur Kuzucu, Franco N~C Wong, Sunao Kurimura, and Sergey Tovstonog.
\newblock {Time-resolved single-photon detection by femtosecond upconversion}.
\newblock {\em Optics Letters}, 33(19):2257, 2008.

\bibitem{Allgaier2017}
Markus Allgaier, Gesche Vigh, Vahid Ansari, Christof Eigner, Viktor Quiring,
  Raimund Ricken, Benjamin Brecht, and Christine Silberhorn.
\newblock {Fast time-domain measurements on telecom single photons}.
\newblock {\em Quantum Science and Technology}, 2(3):034012, sep 2017.

\bibitem{Allgaier2017b}
Markus Allgaier, Vahid Ansari, Linda Sansoni, Christof Eigner, Viktor Quiring,
  Raimund Ricken, Georg Harder, Benjamin Brecht, and Christine Silberhorn.
\newblock {Highly efficient frequency conversion with bandwidth compression of
  quantum light}.
\newblock {\em Nature Communications}, 8:14288, 2017.

\bibitem{Lim2013}
Hwan~Hong Lim, Sunao Kurimura, Toshio Katagai, and Ichiro Shoji.
\newblock {Temperature-dependent sellmeier equation for refractive index of 1.0
  mol\% Mg-doped stoichiometric lithium tantalate}.
\newblock {\em Japanese Journal of Applied Physics}, 52(3 PART 1):0--4, 2013.

\bibitem{Kamada2008}
H.~Kamada, M.~Asobe, T.~Honjo, H.~Takesue, Y.~Tokura, Y.~Nishida, O.~Tadanaga,
  and H.~Miyazawa.
\newblock {Efficient and low-noise single-photon detection in 1550 nm
  communication band by frequency upconversion in periodically poled LiNbO3
  waveguides}.
\newblock {\em Optics Letters}, 33(7):639, 2008.

\bibitem{Shentu2013}
Guo-Liang Shentu, Jason~S Pelc, Xiao-Dong Wang, Qi-Chao Sun, Ming-Yang Zheng,
  M~M Fejer, Qiang Zhang, and Jian-Wei Pan.
\newblock {Ultralow noise up-conversion detector and spectrometer for the
  telecom band}.
\newblock {\em Optics Express}, 21(12):13986--13991, 2013.

\bibitem{Liang2019}
Long~Yue Liang, Jun~Sheng Liang, Quan Yao, Ming~Yang Zheng, Xiu~Ping Xie, Hong
  Liu, Qiang Zhang, and Jian~Wei Pan.
\newblock {Compact all-fiber polarization-independent up-conversion
  single-photon detector}.
\newblock {\em Optics Communications}, 441(February):185--189, 2019.

\bibitem{Zhang2020}
Kong Zhang, Jun He, and Junmin Wang.
\newblock {Two-way single-photon-level frequency conversion between 852 nm and
  1560 nm for connecting cesium D2 line with the telecom C-band}.
\newblock {\em Optics Express}, 28(19):27785, 2020.

\bibitem{Jiang2017}
Wen-Hao Jiang, Jian-Hong Liu, Yin Liu, Ge~Jin, Jun Zhang, and Jian-Wei Pan.
\newblock {125 GHz sine wave gating InGaAs/InP single-photon detector with a
  monolithically integrated readout circuit}.
\newblock {\em Optics Letters}, 42(24):5090, 2017.

\bibitem{Yu2002}
Nan~Ei Yu, Jung~Hoon Ro, and Myoungsik Cha.
\newblock {Broadband quasi-phase-matched second-harmonic generation in
  MgO-doped periodically poled LiNbO3 at the communications band}.
\newblock {\em Optics Letters}, 27(12):1046--1048, 2002.

\bibitem{Yu2003}
Nan~Ei Yu, Sunao Kurimura, Kenji Kitamura, Jung~Hoon Ro, Myoungsik Cha, Satoshi
  Ashihara, Tsutomu Shimura, Kazuo Kuroda, and Takunori Taira.
\newblock {Efficient frequency doubling of a femtosecond pulse with
  simultaneous group-velocity matching and quasi phase matching in periodically
  poled, MgO-doped lithium niobate}.
\newblock {\em Applied Physics Letters}, 82(20):3388--3390, 2003.

\end{thebibliography}

\end{document}